\documentclass{article}
\usepackage{graphicx}

\def\DR{\rm I\kern-1.45pt\rm R}
\def\DC{\kern2pt {\hbox{\sqi I}}\kern-4.2pt\rm C}

\newcommand{\ba}{\begin{array}}
\newcommand{\ea}{\end{array}}
\newcommand{\be}{\begin{equation}}
\newcommand{\ee}{\end{equation}}
\newcommand{\bea}{\begin{eqnarray}}
\newcommand{\eea}{\end{eqnarray}}
\newcommand{\bi}{\begin{itemize}}
\newcommand{\ei}{\end{itemize}}

\usepackage{amscd,amsmath,amssymb}

\begin{document}\begin{center}
{\bf \Large  Antiferromagnetic sawtooth chain with Heisenberg and Ising bonds.}\\
\vspace{0.5 cm} {\large Vadim Ohanyan }
\end{center}
\noindent
{\it Department of Theoretical Physics, Yerevan State University,
Al. Manoogian 1, 0025, Yerevan, Aremnia \\}
{\it Yerevan Physics Institute, Alikhanian Brothers 2, 0036, Yerevan,  Armenia }\\
\\



 {\sl E-mails: ohanyan@yerphi.am}
\begin{abstract}
The sawtooth chain with pairs of $S=1/2$ spins interacting with $XXZ$-interactions placed on each second tooth is considered. All other interaction bonds are taken to be of Ising type. Exact statistical mechanical solution of the model within the direct transfer-matrix technique is obtained. The solution allows one to obtain exact analytic expressions for all thermodynamic functions of the model. Ground stated properties are also investigated, the corresponding ground state phase diagram is presented.
\end{abstract}


\section{Introduction.}
\begin{figure}
  \includegraphics[height=.10\textheight,width=.85\textwidth]{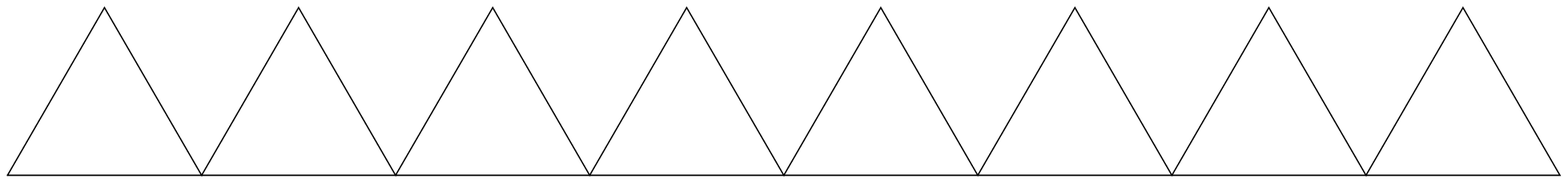}
  \caption{The sawtooth chain.} \label{fig1}
\end{figure}

The sawtooth chain or delta chain is a one-dimensional lattice spin system with a topology of corner-sharing
 triangles (Fig. \ref{fig1}). This system is famous for a number of important features. Physically,  magnetic lattices
 corresponding to the sawtooth chain are found in a number of compounds, in delafossite YCuO$_{2.5}$ \cite{del1, del2} and
 olivines with structures ZnL$_2$S$_4$ (L=Er,Tm,Yb) \cite{oli} to cite a couple of examples. Antiferromagnetic Heisenberg model
 on sawtooth chain is strongly frustrated, however, in contrast to many other frustrated systems, the ground state of sawtooth
 chain is known exactly. Namely, the ground state in case of cyclic boundary conditions is a two--fold degenerated dimerised
 structure where either each left pair of spin at each triangle or each right pair forms spin-singlet
 states (dimers)\cite{kub93}. The excitations of the sawtooth chain are gapped and have topological origin.
 They are found to be "kink" -"antikink"-type domain wall structures \cite{nak96, sen96}.
 Anisotropic variants of sawtooth chain have also been investigated revealing additional features \cite{nak97,blu03}.
 Another important feature of the sawtooth chain as well as some other frustrated spin and Hubbard lattices revealed quite recently
 is the appearance of localized magnons states (dispersionless excitation bands) \cite{rich04, der07a, der07b, rich08}. The localized
 low energy excitations in sawtooth chain which exist due to frustrated geometry, more precisely, due to the triangular plaquette
 affect the low--temperature thermodynamics. The corresponding eigenstates have been constructed for sawtooth chain in case of
 Heisenberg model in Ref. \cite{rich08} and for Hubbard electrons in Ref. \cite{der07a}. Calculating the ground state degeneracies
  for the localized magnon (or electron) states one can obtain explicit expressions for thermodynamic quantities for low temperatures
   and near saturation field \cite{der07a, rich08}. However, the problem of exact description of whole thermodynamics for sawtooth
    chain as well as for many other interacting quantum spin systems is still an open issue.

 One can mention a formal approach to these problem which is not justified properly yet but in some particular cases
 demonstrates rather good agreement with experimental data and numerical calculations. The approximation consists in replacement of some
 or all interaction bonds with Ising ones \cite{oha03}--\cite{per09}. As a result one can obtain an interacting spin system which allows
 one to calculate all thermodynamic functions analytically. In this
 paper we use this approach for exact solution of the system with
 Ising and Heisenberg interaction bonds on sawtooth chain. To
 make possible exact calculation of the partition function within
 transfer-matrix method we consider all interaction bonds of
 sawtooth chain as Ising ones except left bonds on each second
 triangle. Spins connected by these bonds interact by Heisenberg
 $XXZ$ interaction. Such a construction allows one to represent the
 Hamiltonian in the form of sum of block Hamiltonian commutating to
 each other, thus, the exponent in partition function can be
 expanded to the product of exponents corresponding to each block.
 Due to this fact, the method of transfer-matrix can be exploited
 leading to exact calculation of all thermodynamic functions. In the
 first section we formulate the model and give its exact solution,
 then, in the next section, we analyze various ground states of the
 system and draw a corresponding phase diagram. The last section
 contains concluding remarks.


\section{The model and its exact solution.}
\begin{figure}
  \includegraphics[height=.13\textheight,width=.85\textwidth]{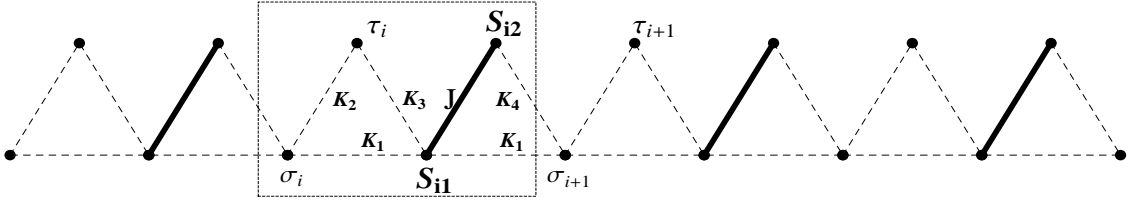}
  \caption{The sawtooth chain with Heisenberg bond on every second triangle. The Heisenberg bond is marked with bold line. Group of spins included into the dotted rectangle correspond to one block with Hamiltonian $\mathcal{H}_i$ }. \label{fig2}
\end{figure}

Let us consider the sawtooth chain in which the left pair of spins on eevery second triangle interacts with the $XXZ$ - Heisenberg interaction, while all other interaction bonds are of Ising type, i.e. the interactions include only $z$-component of spin operators. The corresponding Hamiltonian is suitable to write as a sum of block Hamiltonians where each block contains 5 spins: a pair of spins connected with quantum bond, two adjacent spins from the basement and one spin at the top of the triangle from the left hand side (See Fig. \ref{fig2}):
\begin{eqnarray}
&&\mathcal{H}=\sum_{i=1}^N \left(\mathcal{H}_i+K_2 \sigma_i \tau_i -H_1 \tau_i -\frac{H_2}{2}\left( \sigma_i + \sigma_{i+1}\right)  \right), \label{ham} \\
&&\mathcal{H}_i=J\left(\Delta \left( S_{i1}^x S_{i2}^x+S_{i1}^y S_{i2}^y\right)+S_{i1}^z S_{i2}^z \right)+K_1 S_{i1}^z \left(\sigma_i+\sigma_{i+1} \right)+K_3 S_{i1}^z \tau_i +K_4 S_{i2}^z \sigma_{i +1}-H (S_{i1}^z+S_{i2}^z ), \nonumber
\end{eqnarray}
where $S_{i,a}^{\alpha}$, $a=1,2$, $\alpha =x,y,z$ are $S=1/2$ spin operator components of the spin connected by quantum bond from $i-th$ block, $\sigma_i$ and $\sigma_{i+1}$ are two adjacent spins from the basement and $\tau_i$ is the spin on the top of left triangle; as only $z$-components of all $\sigma$ and $\tau$ spins are included in the interaction, one can consider them just as a classical variable, taking values $\pm 1/2$. Here we also choose the most general form of interaction with different coupling constants $J, K_1, K_2, K_3, K_4 $ for different kind of bonds and different $g$-factors for Zeeman term corresponding to different kind of spins. The partition function of the system then reads:
\begin{eqnarray}
\mathcal{Z}=\sum_{\left(\sigma \right)} \sum_{\left( \tau \right)} \mbox{Sp}_{\mathbf{S}}e ^{-\beta \mathcal{H}}. \label{Z}
\end{eqnarray}
Here one should sum over all values of classical variables $\sigma$ and $\tau$ and take a trace over all states of spin operators $\mathbf{S}$. Because of commutativity of Hamiltonians for different blocks one can expand the exponent and get the product of the terms corresponding to different blocks. After that the traces for each blocks can be taken separately:
\begin{eqnarray}
\mathcal{Z}=\sum_{\left(\sigma \right)}\prod_{i=1}^N\sum_{\tau_i = \pm1/2}\Omega\left(\sigma_i, \sigma_{i+1}| \tau_i \right)e^{\beta H_1 \tau_i - \beta K_2 \tau_i \sigma_i + \beta \frac{H_2}{2} \left(\sigma_i + \sigma_{i+1} \right)}, \label {Z2}
\end{eqnarray}
where
\begin{eqnarray}
\Omega\left(\sigma_i, \sigma_{i+1}| \tau_i \right)=\mbox{Sp}e^{-\beta \mathcal{H}_i}=\sum_{n=1}^4 e^{-\beta \lambda_n \left(\sigma_i, \sigma_{i+1}| \tau_i \right)}, \label{omega}
\end{eqnarray}
and $\lambda_n$ are four eigenvalues of the $\mathcal{H}_i$:
\begin{eqnarray}
&&\lambda_{1,2} \left(\sigma_i, \sigma_{i+1}| \tau_i \right)=\frac{1}{4}J \mp\frac{1}{2}\left(2H-K_1 \left(\sigma_i+\sigma_{i+1} \right)-K_3 \tau_i -K_4 \sigma_{i+1} \right), \label{lam} \\
&&\lambda_{3,4} \left(\sigma_i, \sigma_{i+1}| \tau_i \right)=-\frac{1}{4}J \mp\frac{1}{2} \sqrt{\left( K_1 \left(\sigma_i + \sigma_{i+1} \right)+K_3 \tau_i - K_4 \sigma_{i+1}\right)^2+J^2 \Delta^2}. \nonumber
\end{eqnarray}
In Eq. (\ref{Z2}) the sum over all states of spin $\tau$ in each block can be taken independently from the other, then yielding
\begin{eqnarray}
&&\mathcal{Z}=\sum_{\left(\sigma \right)}\prod_{i=1}^N Z \left( \sigma_i, \sigma_{i+1}\right)e^{\beta \frac{H_2}{2}\left( \sigma_i+\sigma_{i+1}\right)}, \label{Z3}  \\
&&Z \left( \sigma_i, \sigma_{i+1}\right)=\Omega\left(\sigma_i, \sigma_{i+1}| 1/2 \right)e^{\beta \frac{1}{2}\left(H_1-K_2 \sigma_i \right)}+
\Omega\left(\sigma_i, \sigma_{i+1}| -1/2 \right)e^{-\beta \frac{1}{2}\left(H_1-K_2 \sigma_i \right)}.\nonumber
\end{eqnarray}
Thus, the partition function takes the form similar to that of the single chain with $N$ sites with classical variables $\sigma_i$ in them. The partition fuction can be calculated within the standard transfer-matrix technique (See for example \cite{bax}):
\begin{eqnarray}
&&\mathcal{Z}=\sum_{\left(\sigma \right)}\prod_{i=1}^N  T\left(\sigma_i, \sigma_{i+1} \right) = \mbox{Tr} \mathbf{T}^N=\Lambda_+^N+\Lambda_-^N, \label{Z4}
\end{eqnarray}
here the cyclic boundary conditions $\sigma_{i+N} = \sigma_i$ are imposed and 2 by 2 transfer-matrix $\mathbf{T}$ with eigenvalues $\Lambda_{\pm}$ has the following form:
\begin{eqnarray}
{\mathbf{T}}= \left( \begin{array}{lcr}
      Z(1/2,1/2)e^{\beta \frac{H_2}{2}} &
      Z(1/2,-1/2)\\
      Z(-1/2,1/2) &
      Z(-1/2,-1/2)e^{-\beta \frac{H_2}{2}}\label{TM}
      \end{array}
\right).
\end{eqnarray}
Thus, for free energy per one block of the sawtooth lattice with mixed Heisenberg and Ising bonds described above in the thermodynamic limit when only maximal eigenvalue survives from Eq. (\ref{Z4}) one obtains
\begin{eqnarray}
f=-\frac{1}{\beta}\log \left(\frac{1}{2}\left(Z_+e^{\beta \frac{H_2}{2}}+Z_- e^{-\beta \frac{H_2}{2}}+\sqrt{\left( Z_+e^{\beta \frac{H_2}{2}}-Z_- e^{-\beta \frac{H_2}{2}}\right)^2+4 Z_0 \bar{Z}_0} \right) \right), \label{f}
\end{eqnarray}
where the following notation are introduced: $Z_+=Z(1/2,1/2)$, $Z_-=Z(-1/2,-1/2)$, $Z_0=Z(1/2,-1/2)$ and $\bar{Z}_0=Z(-1/2,1/2)$.
Then, one can obtain analytic expressions for all thermodynamic quantities of the systems by taking derivatives of Eq. (\ref{f}). Thus, Eqs. (\ref{Z2})-(\ref{f})  exactly solve the problem of thermodynamics of the systems under consideration. Let us finally list the explicit expressions for the entries of transfer-matrix for particular choice of couplings, mentioned in the beginning of the next section, $K_1=J_1$ and $K_2=K_3=K_4=J$ and for the same value of $g$-factor for all spin, which imply $H_1=H_2$.
\begin{eqnarray}
&&Z_+=2\left(e^{\beta \frac{1}{2}\left(H-1/2 J \right)}\left(e^{-\beta \frac{J}{4}} \cosh\left(\beta \left(H-1/2 (J_1-1/2J) \right) \right)+
e^{\beta \frac{J}{4}} \cosh\left(\beta 1/2 \sqrt{J_1^2+J^2 \Delta^2} \right)\right) \right. \label{ZZ} \\
&&\left.
+ e^{-\beta \frac{1}{2}\left(H-1/2 J \right)}\left(e^{-\beta \frac{J}{4}} \cosh\left(\beta \left(H-1/2 (J_1-3/2J) \right) \right)+
e^{\beta \frac{J}{4}} \cosh\left(\beta 1/2 \sqrt{(J_1-J)^2+J^2 \Delta^2} \right)\right)
 \right), \nonumber \\
 &&Z_-=2\left(e^{\beta \frac{1}{2}\left(H+1/2 J \right)}\left(e^{-\beta \frac{J}{4}} \cosh\left(\beta \left(H+1/2 (J_1-1/2J) \right) \right)+
e^{\beta \frac{J}{4}} \cosh\left(\beta 1/2 \sqrt{(J_1-J)^2+J^2 \Delta^2} \right)\right) \right. \nonumber \\
&&\left.
+ e^{-\beta \frac{1}{2}\left(H+1/2 J \right)}\left(e^{-\beta \frac{J}{4}} \cosh\left(\beta \left(H+1/2 (J_1+3/2J) \right) \right)+
e^{\beta \frac{J}{4}} \cosh\left(\beta 1/2 \sqrt{J_1^2+J^2 \Delta^2} \right)\right)
 \right), \nonumber \\
 &&Z_0=2\left(e^{\beta \frac{1}{2}\left(H-1/2 J \right)}\left(e^{-\beta \frac{J}{4}} \cosh\left(\beta \left(H-1/4J) \right) \right)+
e^{\beta \frac{J}{4}} \cosh\left(\beta 1/2 J\sqrt{1+ \Delta^2} \right)\right) \right. \nonumber \\
&&\left.
+ e^{-\beta \frac{1}{2}\left(H-1/2 J \right)}\left(e^{-\beta \frac{J}{4}} \cosh\left(\beta \left(H+3/4J \right) \right)+
e^{\beta \frac{J}{4}} \cosh\left(\beta 1/2 J \Delta \right)\right)
 \right), \nonumber \\
  &&\bar{Z}_0=2\left(e^{\beta \frac{1}{2}\left(H+1/2 J \right)}\left(e^{-\beta \frac{J}{4}} \cosh\left(\beta \left(H-3/4J) \right) \right)+
e^{\beta \frac{J}{4}} \cosh\left(\beta 1/2 J\Delta \right)\right) \right. \nonumber \\
&&\left.
+ e^{-\beta \frac{1}{2}\left(H+1/2 J \right)}\left(e^{-\beta \frac{J}{4}} \cosh\left(\beta \left(H+1/4J \right) \right)+
e^{\beta \frac{J}{4}} \cosh\left(\beta 1/2 J\sqrt{1+ \Delta^2} \right)\right)
 \right). \nonumber
\end{eqnarray}


\section{Ground states phase diagram for antiferromagnetic couplings}
\begin{figure}
  \includegraphics[height=.13\textheight,width=.85\textwidth]{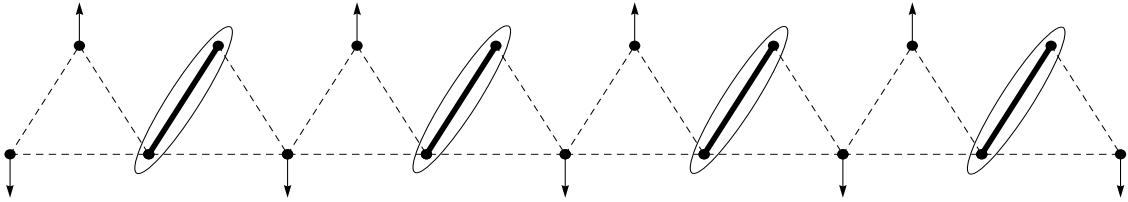}
  \caption{Spin configuration corresponding to the $|AM_-\rangle\ $ antiferromagnetic state. Ovals mark the pairs of spin connected with Heisenberg bonds in  $|S^z=0, -\rangle $ state. One should inverse all arrows to obtain $|S^z=0, +\rangle $ state. } \label{fig3}
\end{figure}

\begin{figure}
  \includegraphics[height=.13\textheight,width=.85\textwidth]{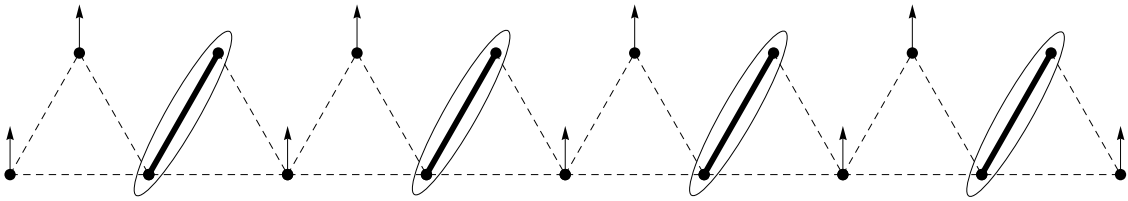}
  \caption{Spin configuration corresponding to the $|F1\rangle\ $ ferrimagnetic state.} \label{fig4}
\end{figure}

Let us analyze different ground states of the model. For the sake of simplicity we restrict ourselves to the case of only two different coupling constants, one along the basement $J_1$ and another one between spins on the top and spins on the basement $J$. This means one should put $K_1=J_1$ and $K_2=K_3=K_4=J$. We also consider only antiferromagnetic coupling $J>0,J_1>0$, the rest cases are to some extent trivial ones. Solving the eigenvalues and eigenvectors problem for block Hamiltonian $\mathcal{H}_i$ and taking into account all combinations of spins $\sigma_i, \sigma_{i+1}$ and $\tau_i$ one can get the following periodic eigenstates for whole chain with period equal to the period of the lattice up to the inversion of all spins: three antiferromagnetic states with $M=0$, two of which are degenerated and differ from each other by the flip of all $\sigma$ and $\tau$ spins(the corresponding energies per one block are presented)
\begin{eqnarray}
&&|AM_+\rangle=\prod_{i=1}^N|S^z=0, +\rangle_i\bigotimes|\sigma_i=\uparrow, \tau_i=\downarrow\rangle \\ \nonumber
&&|AM_-\rangle\prod_{i=1}^N=|S^z=0,-\rangle_i\bigotimes|\sigma_i=\downarrow, \tau_i=\uparrow\rangle, \\ \nonumber
&&\epsilon_{AM}=-\frac{1}{2}\left(J+\sqrt{\left(J-J_1 \right)^2+J^2\Delta^2} \right), \label{AF}
\end{eqnarray}
where $|S^z=0,\pm\rangle_i$ stands for the following state of two $\mathbf{S}$ spins form $i$-th block with total projection equal to zero
\begin{eqnarray}
&&|S^z=0, \pm\rangle=\frac{1}{\sqrt{1+\gamma^2_{\pm}}}\left(|\uparrow\downarrow\rangle-\gamma_{\pm}|\downarrow\uparrow\rangle \right),\label{gamma} \\
&&\gamma_{\pm}=\frac{\pm\left(J-J_1 \right)+\sqrt{\left(J- J_1 \right)^2+J^2 \Delta^2}}{J \Delta}, \nonumber
\end{eqnarray}
This eigenstate corresponds to the $\lambda_3$ eigenvalue from Eq. (\ref{lam}). For arbitrary values of $\sigma$ and $\tau$ surrounding pairs of $\mathbf{S}$ spins coefficient $\gamma$ from Eq. (\ref{gamma}) is
\begin{eqnarray}
\gamma\left(\sigma_i, \sigma_{i+1}| \tau_i \right)=\frac{J\left(\sigma_{i+1}-\tau_i\right)-J_1\left(\sigma_i+\sigma_{i+1}\right)
+\sqrt{\left(J\left(\sigma_{i+1}-\tau_i\right)-J_1\left(\sigma_i+\sigma_{i+1}\right)\right)^2+J^2 \Delta^2}}{J \Delta},\label{gamma2}
\end{eqnarray}
Thus, in these configurations the pair of spins connected with the Ising bonds are aligned opposite to each other, while each pair of spins connected with the Heisenberg bond is in the special state with $S_{tot}^z=0$, see Fig. (\ref{fig3}). Another antiferromagnetic ground state consists of triplet spin configurations on each Heisenberg bond and pairs of spins connected with Ising bond both pointed opposite to the direction of $S=1$ pairs. However, this ground state never realize at $T=0$ for antiferromagnetic region of values of $J$ and $J_1$.
There are also three ferrimagnetic state of the chain when external magnetic field is turned on. All these states have spatial period corresponding to the period of the chain and magnetization equal to 1/2. In the first ferrimagnetic state all pairs of $\mathbf{S}$ spins are in a singlet state, while the rest $\sigma$ and $\tau$ spins are pointed along the field(Fig. \ref{fig4}):
\begin{eqnarray}
|F1\rangle=\prod_{i=1}^N |S^z=0\rangle_i\bigotimes|\sigma_i=\uparrow, \tau_i=\uparrow\rangle, \quad \epsilon_{F1}=-\frac{1}{2}\sqrt{J_1^2+J^2 \Delta^2}-H,\label{F1}
\end{eqnarray}
where eigenstate $|S^z=0\rangle_i$ is given by the same formula as Eq. (\ref{gamma}) but with another value of the coefficient gamma, which is given by Eq. (\ref{gamma2})
\begin{eqnarray}
\gamma=\frac{-J_1+\sqrt{J_1^2+J^2\Delta^2}}{J \Delta}. \label{gamma3}
\end{eqnarray}

\begin{figure}
  \includegraphics[height=.13\textheight,width=.85\textwidth]{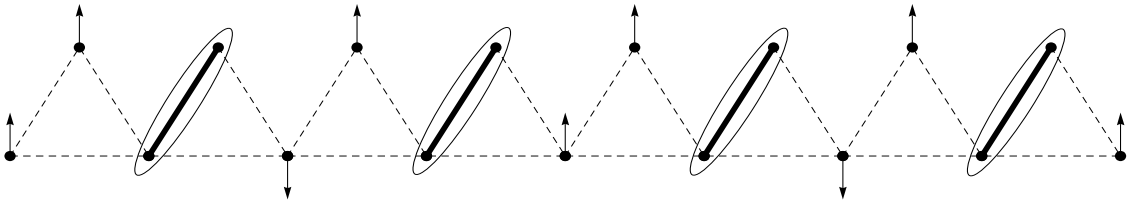}
  \caption{Spin configuration for corresponding to $|SM\rangle$ phase }. \label{fig5}
\end{figure}

\begin{figure}
  \includegraphics[height=.33\textheight,width=.65\textwidth]{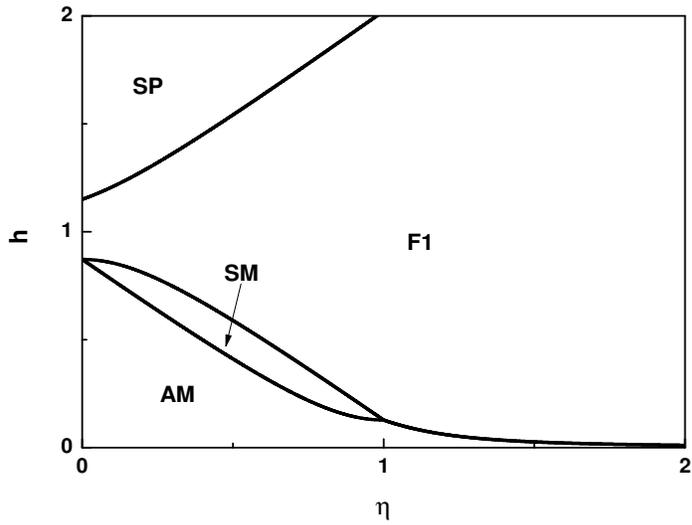}
  \caption{$T=0$ ground state phase diagram for the sawtooth chain with Ising and Heisenberg bonds for $J>0$, $J_1>0$ in ($\eta, h$)-plane for $\Delta=0.3$.  }. \label{fig6}
\end{figure}

In the rest two ferrimagnetic states all pairs of $\mathbf{S}$ spins are in the triplet polarized state with $S^z=1$, while  $\sigma$ and $\tau$ spins surrounding them are pointed opposite to each other:
\begin{eqnarray}
&&|F2\rangle=\prod_{i=1}^N |\uparrow\uparrow\rangle_i\bigotimes|\sigma_i=\uparrow, \tau_i=\downarrow\rangle, \quad \epsilon_{F2}=\frac{1}{4}J_1-H \label{F1} \\
&&|F3\rangle=\prod_{i=1}^N |\uparrow\uparrow\rangle_i\bigotimes|\sigma_i=\downarrow, \tau_i=\uparrow\rangle, \quad \epsilon_{F3}=-\frac{1}{4}J_1-H. \nonumber
\end{eqnarray}
And finally in the spin polarized phase all spins are pointed along the field,
\begin{eqnarray}
|SP\rangle=\prod_{i=1}^N |\uparrow\uparrow\rangle_i\bigotimes|\sigma_i=\uparrow, \tau_i=\uparrow\rangle, \quad \epsilon_{SP}=J+\frac{1}{2}J_1-2H. \label{SP}
\end{eqnarray}
Besides the states with non-broken lattice symmetry the system under consideration can posses another ground state with a spatial period equal to two blocks. This state can be realized at $T=0$ only for $\eta=\frac{J_1}{J} \leq 1$. This state can be obtained from $|AM_-\rangle$ by flipping every second $\sigma$ spin in the basement of the chain (Fig. \ref{fig5}). This state with broken translational symmetry corresponds to magnetization of the system equal to 1/4. We will refer to it as a spin modulated phase, $|SM\rangle$. The spacial period of the state and value of magnetization per site are in full agreement with Oshikawa--Yamanaka-Affleck criterion \cite{OYA}. The corresponding energy per one block is
\begin{eqnarray}
\epsilon_{SM}=-\frac{1}{4}J \left(1+|\Delta|+\sqrt{1+\Delta^2} \right)-\frac{1}{2}H. \label{SM}
\end{eqnarray}
Now it is straightforward to draw the $T=0$ ground state phase diagram for antiferromagnetic couplings $J>0$, $J_1>0$. In Fig. (\ref{fig6}) one can see the phase boundaries in $(\eta, h)$ plane ($h=H/J$) for $\Delta=0.3$. An interesting feature of the system is that at antiferromagnetic values of couplings the ground state for $H=0$ is unaffected by the value of $\eta$. Thus, the unique ground state is either $|AM_-\rangle$ or $|AM_+\rangle$. One should distinguish two regions of $\eta$, namely,  for $\eta<1$ increasing the magnetic field one can see transition to spin modulated phase with $M=1/4$ prior to appearance of the ferrimagnetic structure $F1$ with $M=1/2$. However, for $\eta>1$ antiferromagnetic phase changes immediately with the ferrimagnetic one, thought width of the $AM$ phase region (the value of the $h$ at which the zero temperature quantum phase transition takes the place) with  $\eta \to \infty$ asymptotically goes to zero. So, one can see a small closed region of $SM$ phase in the  $(\eta, h)$ plane. The values $\eta=1, h=0$ the system demonstrate additional feature, the macroscopic degeneracy of the ground state. At this point the the $\mathbf{S}$ spins in each block are in the singlet state, whereas its adjacent $\sigma$ and $\tau$ can take arbitrary value with one restriction $\sigma+\tau=0$. Thus, in each block $\sigma$ and $\tau$ can freely take either $\sigma=1/2, \tau=-1/2$ or $\sigma=-1/2, \tau=1/2$ values. Thus, one obtains macroscopic two-fold degeneracy at $J=J_1$. One can refer to this state as a frustrated antiferromagnetic state:
\begin{eqnarray}
|FR \rangle=\prod_{i=1}^N|S^z=0\rangle_i\bigotimes|\xi_i\rangle, \label{FR}
\end{eqnarray}
where $|\xi_i\rangle$ can be either $|\sigma_i=\uparrow, \tau_i=\downarrow\rangle$ or $|\sigma_i=\downarrow, \tau_i=\uparrow\rangle$
And finally the equations of phase boundaries between four phases presented in the phase diagram in Fig. (\ref{fig5}) are listed below.
\begin{eqnarray}
&&\mbox{between} \quad |AM\rangle \quad \mbox{and} \quad |SM\rangle :\quad h=\frac{1}{2}\left(1-|\Delta|-\sqrt{1+\Delta^2} \right)+\sqrt{\left(1-\eta\right)^2+\Delta^2 },\label{b} \\
&&\mbox{between} \quad |AM\rangle \quad \mbox{and} \quad |F1\rangle :\quad h=\frac{1}{2}\left(1+\sqrt{\left(1-\eta\right)^2+\Delta^2}-\sqrt{\eta^2+\Delta^2} \right), \nonumber \\
&&\mbox{between} \quad |SM\rangle \quad \mbox{and} \quad |F1\rangle :\quad h=\frac{1}{2}\left(1+|\Delta|+\sqrt{1+\Delta^2}\right)-\sqrt{\eta^2+\Delta^2} , \nonumber \\
&&\mbox{between} \quad |F1\rangle \quad \mbox{and} \quad |SP\rangle :\quad h=1+\frac{1}{2}\left(\eta+\sqrt{\eta^2+\Delta^2} \right). \nonumber
\end{eqnarray}
\section{Conclusion}

\begin{figure}
  \includegraphics[height=.33\textheight,width=.65\textwidth]{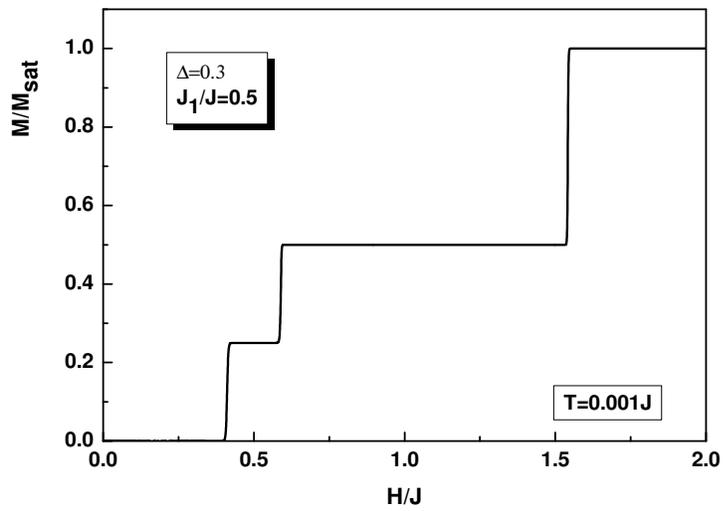}
  \caption{Magnetization process for $\eta=0.5$ at extremely low temperature $T=0.001J$ and $\Delta=0.3$. Three magnetization plateaus at $M=0$, $M=1/4$ and $M=1/2$ correspond to $AM$, $SM$ and $F1$ phases.} \label{fig7}
\end{figure}

\begin{figure}
  \includegraphics[height=.33\textheight,width=.65\textwidth]{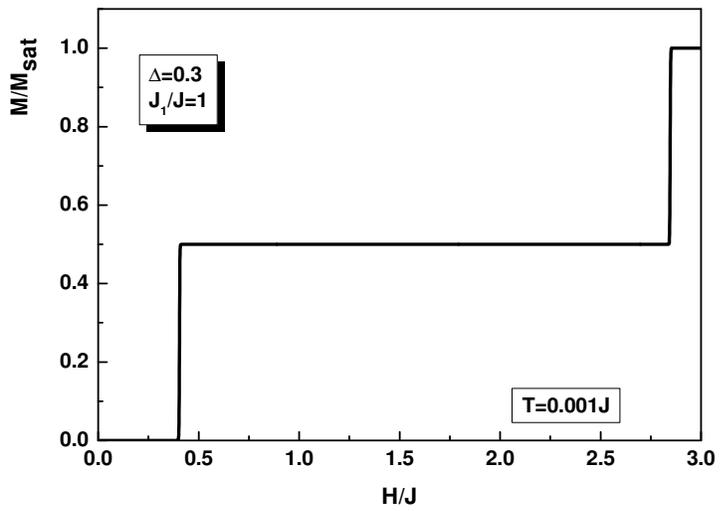}
  \caption{Magnetization process for $\eta=1$ at extremely low temperature $T=0.001J$ and $\Delta=0.3$. Spin modulated phase corresponding to the plateau at $M=1/4$ is absents here.
  One can see only two plateaus at $M=0$ and $M=1/2$ corresponding to $FR$ and $F1$ phases. } \label{fig8}
\end{figure}

In this paper using the example of sawtooth chain we considered the possibility to obtain exact thermodynamic solution
for the one--dimensional quantum spin systems by considering their
counterparts where some interaction bonds are changed with the Ising
ones. Namely, we
considered the pairs on $S=1/2$ quantum spins interacting with $XXZ$
interaction arranged in the sawtooth chain in such a way that allows
one to use transfer-matrix technique for exact calculations. This
result continues the series of investigations of the subject
performed earlier for other one-dimensional spin systems with Ising
and Heisenberg bonds \cite{oha03}-\cite{per09}. These results are
not only of academic interest. There are many one--dimensional spin
systems which are the models of real magnetic material and which are
not integrable. Until now, only the actually reliable way to describe
the thermodynamic and magnetic properties of such a system is
laborious numerical calculations. However, results of recent
investigations are evidence that changing some bonds of the
system with Ising ones, on the one hand, makes the thermodynamical problem
exactly solvable and, on the other hand, does not change drastically
magnetic properties of the system at least in case when only
ferromagnetic bounds have been changed. One can look at the plots of
low--temperature magnetization curves for the system considered in
the paper (Figs. (\ref{fig7}), (\ref{fig8})). As it can be seen from
phase diagram (Fig. \ref{fig6}), there is qualitative difference
between $\eta<0$ and $\eta \geq 1$ cases. For $\eta<1$ one obtains
the curve with three intermediate magnetization plateaus, at $M=0$,
$M=1/4$ and $M=1/2$, which corresponds to the stability regions of
$AM$, $SM$ and $F1$ structures respectively. In Fig. (\ref{fig7})
one can see the corresponding plot for $\Delta=0.3$ and $\eta=0.5$.
For $\eta=1$ and $\Delta=0.3$ one can see in Fig. (\ref{fig8}) only
one plateau at $M=1/2$. This result qualitatively corresponds to
that obtained in purely quantum sawtooth chain \cite{rich08}.
Numerically obtained magnetization curves for sawtooth Heisenberg
chain contain only plateaus at $M=0$ and $M=1/2$ for whole range of
parameters. Though, there is rather large discrepancy in
quantitative characteristics, such as positions of the terminal points
of the plateau, but this can be explained by the fact that in our
case we changed antiferromagnetic bonds with Ising ones, which
generally speaking makes much more changes into the thermodynamics
of the system than changing ferromagnetic bonds. An additional feature
which is absent in case of purely quantum antiferromagnetic sawtooth
chain is the plateau at $M=1/4$ for $\eta<1$. In order to obtain
more satisfactory correspondence between numerical results for
purely quantum system and its mixed counterpart,one can consider a
larger quantum cluster, containing three or more spins.



\section{Acknowledgments}
  This work was partly supported by the grants CRDF-UCEP - 06/07 and ANSEF-1586-PS. The author would like to
  thank Oleg Derzhko for useful discussions and Lev Ananikian and Diana Antonosyan for help in preparing the figures.


\newpage




%

\end{document}